\documentclass[preprint]{aastex}

\newcommand{\simle}{\mbox{$\stackrel{<}{_{\sim}}$}}

\slugcomment{Accepted in Astrophysical Journal}

\shorttitle{Bipolar Nebula CIT 6}
\shortauthors{Monnier et al.}

\begin{document}

\title{Diffraction-limited near-IR
imaging at Keck reveals asymmetric, time-variable
nebula around carbon star CIT 6}


\author{J. D. Monnier\altaffilmark{1}, P. G. Tuthill\altaffilmark{2}
and W. C. Danchi\altaffilmark{3,4}}

\altaffiltext{1}{Smithsonian Astrophysical Observatory MS\#42,
60 Garden Street, Cambridge, MA, 02138}
\altaffiltext{2}{Chatterton Astronomy Dept, 
School of Physics, University of Sydney, NSW 2006, Australia}
\altaffiltext{3}{NASA Goddard Space Flight Center,
Infrared Astrophysics, Code 685, Greenbelt, MD 20771}
\altaffiltext{4}{Space Sciences Laboratory, University of California, Berkeley,
Berkeley, CA  94720-7450}
\email{jmonnier@cfa.harvard.edu, gekko@physics.usyd.edu.au, wcd@ssl.berkeley.edu}


\begin{abstract}
We present multi-epoch, diffraction-limited images of the nebula
around the carbon star CIT~6 at 2.2\micron~and 3.1\micron~from
aperture masking on the Keck-I telescope.  The near-IR nebula is
resolved into two main components, an elongated, bright feature showing
time-variable asymmetry and a fainter component about
60\,milliarcseconds away with a cooler color temperature.  These
images were precisely registered ($\sim$35~milliarcseconds) with
respect to recent visible images from the Hubble Space Telescope
\citep{trammell2000}, which showed a bipolar
structure in scattered light.  The dominant near-IR feature
is associated with the northern lobe of this scattering nebula, and
the multi-wavelength dataset can be understood in terms of a
bipolar dust shell around CIT~6.  Variability of the near-IR morphology
is qualitatively 
consistent with previously observed changes in red polarization, 
caused by varying illumination geometry due to non-uniform dust production.
The blue emission morphology and
polarization properties can not be explained by the above model
alone, but
require the presence of a wide binary companion in the vicinity of
the southern polar lobe.  The physical mechanisms responsible for the
breaking of spherical symmetry around extreme carbon stars, such as
CIT~6 and IRC+10216, remain uncertain.

\end{abstract}

\keywords{techniques: interferometric,
stars: AGB and post-AGB, stars: circumstellar matter,
stars: winds}


\section{Introduction}
As evidenced by its primary designation, CIT~6 (= RW LMi = GL 1403 =
IRC +30219 = IRAS~10131+3049) was discovered as a bright infrared
source without a known visible counterpart at the time of the Caltech infrared
survey \citep{ulrich66}.  The identification of CIT~6 as a carbon star 
is based on its
near-infrared CN bands \citep{wisniewski67,gaustad69} and its thick
carbon-rich dust shell, responsible for the heavy visible obscuration
\citep[e.g.,][]{strecker74}. Monitoring of the infrared variability, 
first recognized
by \citet{ulrich66}, has led to an established period of
$\sim$628~days for this long-period variable
\citep{alksnis75,alksnis95,taranova99}.  Based on the pulsational period, the
distance to CIT~6 has been estimated by \citet{cohen96} as 
400$\pm$50\,pc from application of
the period-luminosity relations established for O-rich Miras (LMC and
galactic) and C-rich Miras (LMC only).

Both the visible and infrared flux from CIT~6 is highly polarized,
indicating an asymmetric distribution of circumstellar material
\citep{kruszewski68, dyck71}.  In addition, the fractional polarization
is variable \citep{kruszewski71, dyck71, kruszewski73, alksnis75} and
the polarization angle is currently strongly wavelength-dependent,
changing $\sim$90\arcdeg~between the blue and the red
\citep{kruszewski73,cohen82,trammell94}.  This distinctive behavior is
likely linked to the excess UV/blue flux and unusual emission lines
which have been observed, perhaps due to the presence of an unseen
(hot) companion \citep{alksnis88} or circumstellar shocks
\citep{cohen80,cohen82}.

Although no previous high resolution imaging of this source has been
reported, the characteristic sizes of the near- and mid-infrared
emission regions have been measured.  \citet{dyck84} used speckle
interferometry to estimate the 2.2\micron, 3.5\micron, and 
4.8\micron~sizes to be $\sim$70\,mas, 100\,mas, and 144\,mas respectively 
(Gaussian FWHM).  In addition, \citet{lipman98} reported a 11\micron~
FWHM of 150\,mas using the Infrared Spatial Interferometer.

Recently developed observing techniques, from both the ground and space, 
allow the asymmetries of the dust envelope to be observed directly. 
In this paper we report multi-epoch, 
diffraction-limited observations of the circumstellar
dust shell at 1.6\micron, 2.2\micron, and 3.1\micron~from the Keck-I 
telescope.  Using 
aperture masking interferometry, we attain similar spatial
resolution in the infrared 
to that of the Hubble Space Telescope in the visible,
and the brightness distributions at these different wavelengths 
have been registered and
compared 
in order to probe the circumstellar environment of this enigmatic system.

\section{Observations}
\subsection{Infrared Interferometry with Keck-I}
Aperture masking interferometry was performed by placing aluminum
masks in front of the Keck-I infrared secondary mirror.  This
technique converts the 10-m primary mirror into an
interferometric array, allowing the Fourier amplitudes and closure
phases for a range of baselines to be recovered with minimal
``redundancy'' noise \citep[e.g.,][]{baldwin86}.  In this experiment,
both non-redundant and doubly-redundant pupil configurations were
employed.  The Maximum Entropy Method (MEM) \citep{mem83,sivia87} was
used to reconstruct diffraction-limited images from the
interferometric data.  
MEM does produce structure
in the maps beyond the formal diffraction limit, an extrapolation
based on the positivity and finite-support image constraints, as well
as the ``maximum entropy'' regularization scheme.  In some cases, this
structure can be spurious and result in well-known MEM artifacts, such
as ringing of point sources embedded in nebulosity \citep{nn86}.  For
CIT~6, strong (non-zero) closure phase signals suppress these
artifacts and result in a high fidelity map.  
In order to check the reliability of the
reconstructions, MEM results were compared to the results
of the CLEAN reconstruction algorithm \citep{hogbom74}.  
Results were always consistent, with smoothed MEM maps being comparable to the
the results of CLEAN, which does not generate similar super-resolution and 
hence has less fine structure in its images.
Further engineering and performance details may be found 
in \citet{mythesis} and
\citet{tuthill2000a}.

CIT~6 was observed at 6 different epochs between 1997 January and 2000
January, at the Keck-I telescope using the Near-Infrared Camera
\citep{ms94,matthews96} and with a variety of aperture masks.  Important
observing information can be found in Table~\ref{table:journal},
including spectral filter characteristics, the number of speckle
frames, and the unresolved stars used for calibrating the atmosphere
plus telescope transfer function.  In all cases, the integration time
of each frame was 0.137\,s.

\begin{table}[t]
\begin{center}
\caption{Journal of Observations \label{table:journal} }

\begin{tabular}{lccccl}
\tableline\tableline
Date 	& $\lambda$ &	FWHM $\Delta \lambda$& Number of & Calibrator & Aperture Mask\tablenotemark{a} \\
(U.T.)	& ($\mu$m)  &	($\mu$m)	& Frames    & Star       & \\
\tableline
1997 January 30 & 2.260 & 0.053 & 100 & $\mu$ UMa & 15-Hole Golay \\
	        & 2.260 & 0.053 & 200 & $\mu$ UMa & Annulus       \\
		& 3.083 & 0.101 & 100 & $\mu$ UMa & 15-Hole Golay \\
		& 3.083 & 0.101 & 200 & $\mu$ UMa & Annulus       \\
1997 December 17& 1.658 & 0.333 & 100 & HD 83362 & Annulus \\
		& 2.260 & 0.053 & 100 & $\mu$ Leo & Annulus \\
		& 2.269 & 0.155 & 100 & $\pi$ Leo/$\alpha$ Hya & 21-Hole Golay \\
		& 3.083 & 0.101 & 100 & $\mu$ Leo & Annulus \\
		& 3.083 & 0.101 & 200 & $\pi$ Leo/$\alpha$ Hya & 21-Hole Golay \\
1997 December 19& 2.260 & 0.053 & 100 & $\mu$ UMa & 21-Hole Golay \\
		& 3.083 & 0.101 & 100 & $\mu$ UMa & 21-Hole Golay \\
1998 April 15	& 2.269 & 0.155 & 100 & $\mu$ UMa & 21-Hole Golay \\
		& 3.083 & 0.101 & 100 & $\mu$ UMa & 21-Hole Golay \\
1998 April 16 	& 1.658 & 0.333 & 100 & HD 86728 & Annulus \\
		& 2.269 & 0.155 & 100 & HD 86728 & Annulus \\
		& 3.083 & 0.101 & 100 & $\mu$ UMa & Annulus \\
1998 June 5	& 2.269 & 0.155 & 100 & $\mu$ UMa & 21-Hole Golay \\
		& 3.083 & 0.101 & 100 & $\mu$ UMa & 21-Hole Golay \\
1999 February 5 & 2.269 & 0.155 & 100 & $\mu$ UMa & 21-Hole Golay \\
		& 3.083 & 0.101 & 100 & $\mu$ UMa & 21-Hole Golay \\
2000 January 26 & 2.269 & 0.155 & 100 & $\alpha$ Lyn & 21-Hole Golay \\
		& 3.083 & 0.101 & 100 & $\alpha$ Lyn & 21-Hole Golay \\
\tableline
\end{tabular}
\tablenotetext{a}{15- and 21-Hole Golay masks refer to non-redundant geometries of 35-cm
holes with longest baselines of 9.26-m and 9.55-m respectively, as projected onto the 
primary mirror.  The annulus mask is doubly-redundant with a projected inner and outer
radii of 8.22-m and 8.73-m.} 
\end{center}
\end{table}

\begin{figure}
\begin{center}
\includegraphics[height=7.5in]{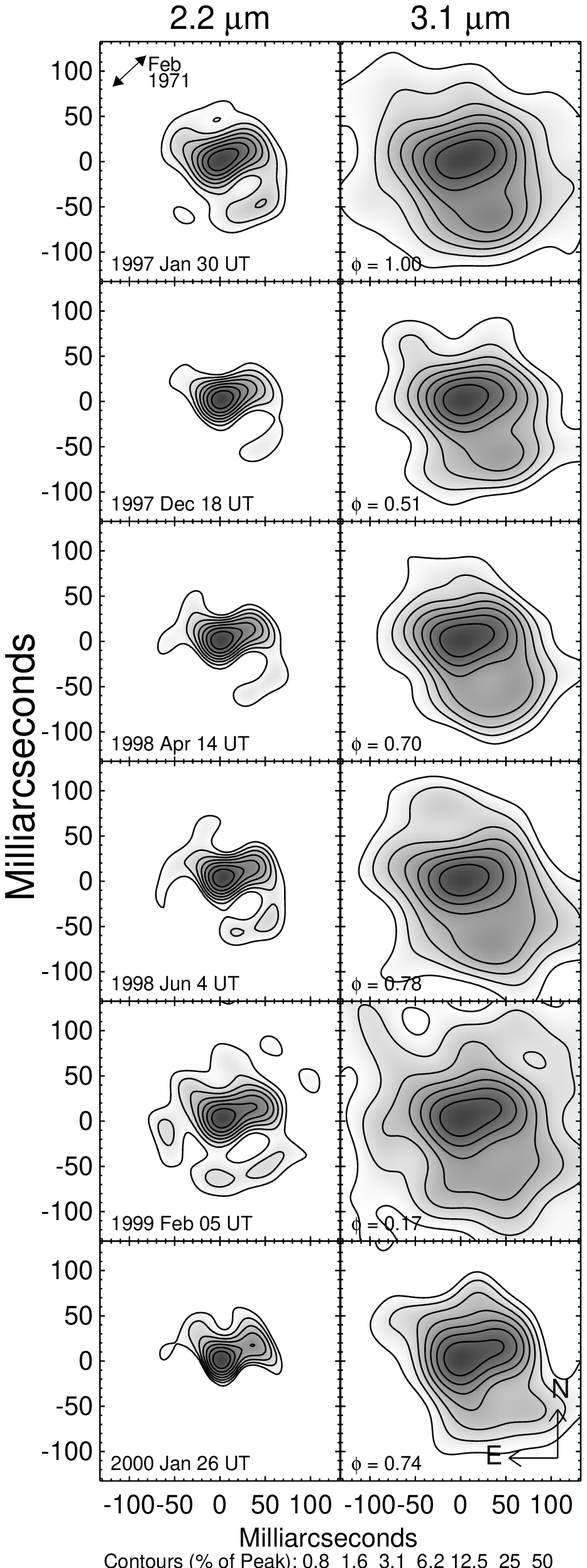} 
\caption{Multi-epoch images of CIT~6 at 2.2\micron
~({\em left panels}) and 3.1\micron ~({\em right panels}). 
Each epoch is labeled by the U.T. date of observation and the
pulsational phase of CIT~6 according to \citet{taranova99}. 
The arrow labeled ``Feb 1971'' in the first panel
represents the most recently published 
polarization direction of the electric field at K band \citep{dyck71}. 
Contour levels are logarithmic, each representing a factor of two, and
 are 0.78\%, 1.56\%, 3.13\%, 6.25\%, 12.5\%, 25\%, and 50\% of the peak.
\label{fig:cit6}}
\end{center}
\end{figure}

Figure\,\ref{fig:cit6} shows MEM reconstructed images of CIT~6 at
2.2\micron~and 3.1\micron~from all epochs.  In cases where multiple
observations exist at a given epoch, images were reconstructed from
the separate datasets and then averaged together, weighted by the
dynamic range of each image.
In addition, the last known
direction of the polarized component of the electric field
in the K band has been included in the first panel for easy reference
\citep{dyck71}.
At both colors the dominant component, appearing at the map center, 
appears as a bright feature, 
elongated to the West-Northwest at a position angle of 
about $290^\circ$.
A second component, visible but faint in the 2.2\,$\mu$m images, and more
prominent at 3.1\,$\mu$m, appears separated by $\sim$60~milliarcseconds
(mas) at a position angle (PA) of $\sim$215\arcdeg from the map center.
These components will be referred to as the northern and southern components
respectively.
These data will be discussed further in the next section.

While CIT~6 was observed near 1.6\micron, low signal-to-noise ratio
prevented accurate image reconstructions.  However, from the
visibility data, it is clear that the dominant component is compact
and elongated along PA $\sim$286\arcdeg, similar to that seen at 2.2
\micron~and 3.1\micron.  Fitting a two-dimensional Gaussian to the visibility
data yielded a major axis of $\sim$24~mas and minor axis of
$\sim$16~mas.  This simple model does not adequately fit the data
($\chi^2$ per degree of freedom $\sim$5), and can only be viewed as a
crude description of the actual brightness distribution.

\subsection{Hubble Space Telescope images}
\begin{figure}[t]
\begin{center}
\includegraphics[width=6in]{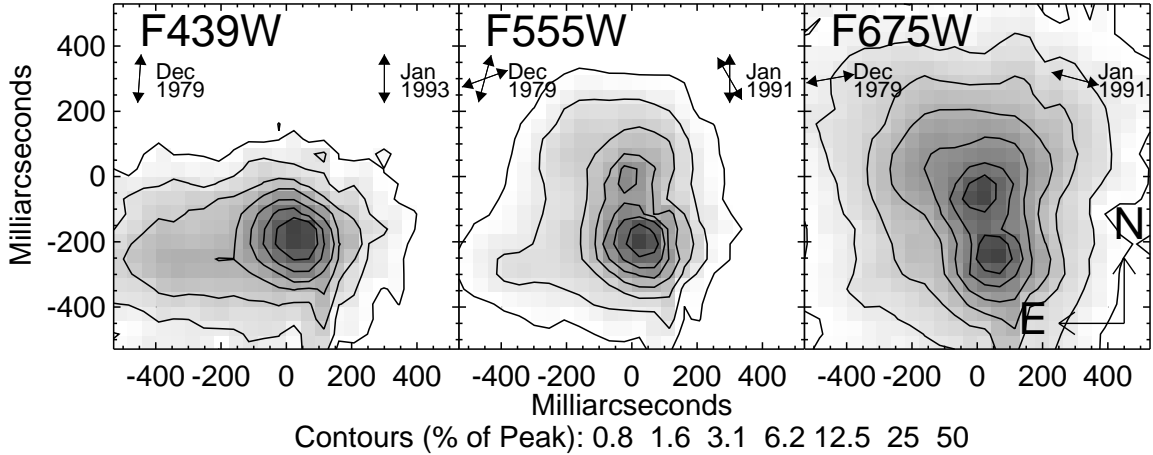} 
\caption{Multi-wavelength images of CIT~6
from the Hubble Space Telescope \citep{trammell2000}, 
using filters F439W (0.429\micron), 
F555W (0.525\micron), and F675W
(0.674\micron).  
Each contour level of intensity represents a factor of two in
surface brightness, and corresponds to 0.78\%, 1.56\%, 3.13\%, 6.25\%, 12.5\%, 25\%, and 50\% of the peak intensity.
The direction of the polarized electric field is shown
in the upper-left and upper-right of each panel, corresponding to two
epochs of observations \citep{cohen82,trammell94}.  Because of significant
wavelength-dependence of the polarization
within the broad bandpass of the F555W filter, 
an additional pair of polarization vectors have been included in 
the middle panel, to
represent the range of angles present in the spectropolarimetric observations.
\label{fig:wfpc2}}
\end{center}
\end{figure}

CIT~6 has also recently been imaged with the Hubble Space Telescope (HST) using
both WFPC2 and NICMOS instruments.  
The WFPC2 observations were carried out on 
1996 April 2, and the reduced images discussed here were 
provided by S. Trammell in advance of publication \citep{trammell2000}. 
The plate scale of the original images was 46~mas per pixel.
Figure~\ref{fig:wfpc2} shows the logarithmic intensity of CIT~6 using
filters F439W ($\lambda$=0.429\micron, $\Delta\lambda$=0.046\micron),
F555W ($\lambda$=0.525\micron, $\Delta\lambda$=0.122\micron), and
F675W ($\lambda$=0.674\micron, $\Delta\lambda$=0.089\micron).	
See the WFPC2 Instrument Handbook for more information on the filter
bandpasses ($\Delta\lambda$ is roughly the FWHM).
The polarization of the electric field in each of these bandpasses
has been included in the figure, both \citet{cohen82} and 
\citet{trammell94} results.  Because of strong wavelength-dependence
of the polarization direction within the large bandpass of the
F555W filter, two polarization vectors for each epoch have been 
plotted to represent the range of observed angles.

The
reddest image (F675W) consists of two components, or lobes, 
of similar brightness,
separated by $\sim$190~mas along PA 190\arcdeg.  Moving towards
shorter wavelengths, the F555W image still shows evidence for two
lobes, but the northern component is considerably fainter with respect
to the southern lobe.  Lastly, the F439 image shows only the southern
component, exhibiting a peculiar comet-like asymmetry at a position angle
roughly orthogonal to the two lobes seen in red.  Rayleigh scattering
along this PA would explain the different angle of polarization for
the blue light, compared to the red emission \citep[e.g.,][]{cohen82}.
Detailed analyses of these images can be found in \citet{trammell2000}.

CIT~6 was observed multiple times by NICMOS before the cryogen
expired, and the data from 1998 February 23 and 1998 May 5 were used
here.  The calibrated data were retrieved from the HST Data Archive,
including the observation log files.  The images themselves will not
be discussed in this paper, but rather the data will be used for
extracting relative astrometry.  This has allowed the Keck infrared
images to be registered with respect to the WFPC2 images as is
discussed fully in the next section.

\subsection{Image Registration}
\label{section:registration}
Even though the Keck imagery contains no positional information and
HST absolute astrometry is only good to $\sim$1\arcsec, it has been  possible
to link the infrared and visible reference frames to about
35~milliarcseconds.  This is because CIT~6 has a companion star
located only $\sim$10.4\arcsec~away \citep[hereafter referred to as
the reference star]{bastian88}, allowing precise {\em relative}
astrometry.  In this section we describe the methodology for
extracting precise relative astrometry from both WFPC2 and NICMOS
images.

\subsubsection{WFPC2}
Deriving the position of CIT~6 relative to the reference star is
straightforward because both appear on the PC chip in all of the WFPC2
images, which has a nominal 
plate scale of 45.5 mas/pixel.  However, for the
images taken with filters F439W and F555W, the reference star is
slightly saturated, reducing the precision of the centroiding process.
When not saturated, the centroid was defined by fitting a pixelized, 
two-dimension Gaussian profile to a 3x3 pixel box centered on the peak pixel.
When saturated, the centroid was defined by fitting a Gaussian
to the non-saturated pixels contained in a 7x7 pixel box.  Tests
were performed on unsaturated point spread functions and differences in the
estimated centroids resulting from 
these two methods were always less than 6~mas.
The IRAF-STSDAS task {\em metric} was translated into IDL and used 
to correct for the
geometric field distortion \citep{gilmozzi95}, however corrections were always
$\simle$0.3 pixels ($\sim$13.7~mas).  
For F439W only the south
component was visible, but both components of the CIT~6 nebula were
measured relative to the reference source for F555W and F675W images.
Although the HST Data Handbook claims that this process should be
accurate to 5~mas for relative astrometry on the same chip, we have
assigned more pessimistic error estimates because the centroid was not
as well defined for the distorted isophotes of the extended
emission, and because of saturation in the core of the reference star,
most pronounced in F555W.  Relative astrometry with estimates of
uncertainty are included in Table~\ref{table:astrometry} and the
results are plotted in Figure~\ref{fig:astrometry}.

\subsubsection{NICMOS}
NICMOS observations of CIT~6 have been carried out (by other
workers, P.I. Gary Schmidt), 
and reside on the HST Data Archive.  Although the spatial 
resolution of HST is four times worse than the Keck-I Telescope, 
these data are useful in the context of this paper 
for linking the position of the infrared
peak to the visible WFPC2 images.  Geometric field distortion was corrected
according to \citet{cox97}, while the appropriate platescale was
determined from the online calibration file 
(\url{http://www.stsci.edu/instruments/nicmos/NICMOS\_status/platescale.dat}).
When the images were not saturated, the centroid was determined as above,
using a pixelized Gaussian to 
fit to a 3x3 pixel square surrounding the peak pixel.
However, two practical problems complicated the analysis.

Firstly, the field-of-view of the NIC1 chip (used for J and H band
observations) is only 11\arcsec$\times$11\arcsec; not large enough to 
image both CIT~6 and the reference star without
placing CIT~6 on the very edge of the chip.  However, high precision relative
astrometry can be maintained by executing a ``small angle maneuver.'' 
When the telescope pointing is changed using this procedure,
the angular offset between images should be known to
$\sim$10~mas, since the same two guide stars are being used 
(G. Schneider, private communication).
By using the RA\_AVG, DEC\_AVG, and ROLL\_AVG keywords in the 
JIF observing log file, the relative offsets between nearby pointings
can be determined.
We applied this procedure to each exposure of an observation to 
measure the relative separation between
CIT~6 and the reference star in filters F110W and F160W for the
observations of 1998 May 5.  This method was tested using F220M images
taken before and after 
a small angle maneuver, both of which contain the reference
source.  Two epochs of such data exist (see below) and 
the residual
astrometric error was found to be only $\simle$12~mas for these test cases.

A second complication was that 
the F160W and F220M images of CIT~6 were heavily saturated.
In these cases, 
the location of the CIT~6 centroid
was estimated by measuring the intersection point
of the diffraction spikes.  We tested the accuracy of this method by
applying it to the unsaturated F110W image and found an offset of 
only $\simle$13\,mas with respect to the centroid determined by a 
Gaussian fit.  
However, since the diffraction pattern is 
wavelength-dependent, this does not assure such accuracy for longer 
wavelengths.  We have assigned a conservative positional 
uncertainty of 0.5 pixels (22~mas for NIC1, 35~mas for NIC2) 
to measurements made using this method.  

The NIC2 chip has a larger field of view (19.2\arcsec$\times$19.2\arcsec)
which allowed both reference star and CIT~6 to appear on the same chip
for the F220M data.  Hence, the relative astrometry in this waveband
does not suffer from possible offset errors associated with the
small angle maneuver, but only those due to saturation of CIT~6 and
uncertainty in the platescale and field distortion.
As another cross-check to the accuracy, we applied this procedure to
an additional epoch of F220M data from 1998 February 11 
(which had a significantly different roll angle), and the
two epochs agreed well within assigned uncertainty, indicating
that our 35~mas estimate for error in the diffraction-spike centroid method
is indeed quite conservative.
Complete results appear in tabulated and graphical forms in 
Table\,\ref{table:astrometry} and Figure\,\ref{fig:astrometry}.
HST imagery is rarely used to perform relative astrometry at this level of
precision, and it is possible that unknown calibration errors may corrupt
these results.  Since the registration of the infrared and visible images is 
critical for interpreting the emission morphology, we are hopeful that
independent confirmation of these results can be obtained.

\subsubsection{Proper motion of CIT~6}
\label{section:motion}
\begin{figure}
\begin{center}
\includegraphics[angle=90,width=4in]{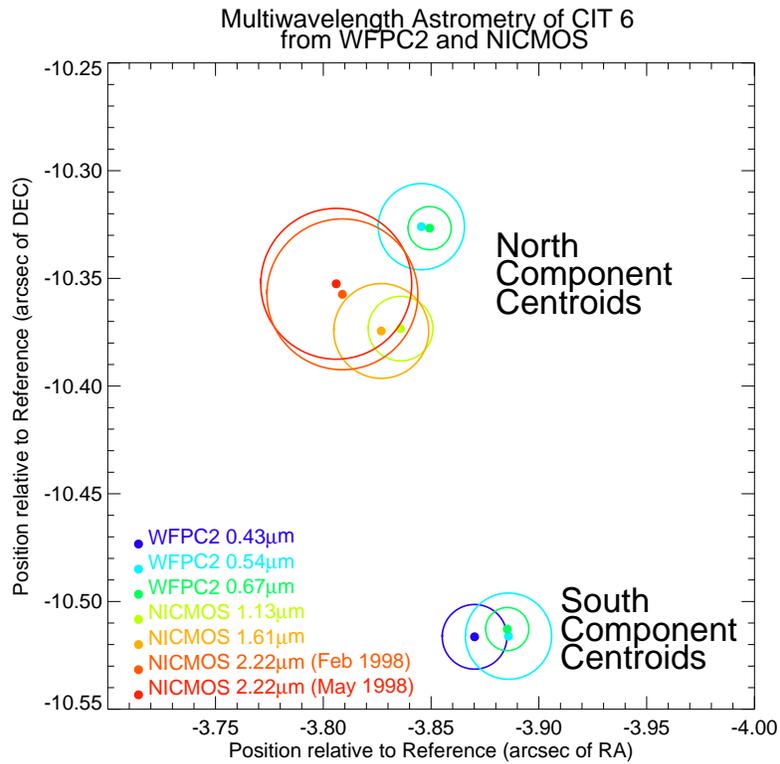} 
      \caption{Multiwavelength astrometry of both components of CIT~6.
The colored circles represent the error circle for the position of
the centroid for each HST image, both WFPC2 and NICMOS.  Note that the
NICMOS centroid is most likely associated with the northern component seen
by WFPC2.  North is up and East is towards the left .
\label{fig:astrometry}}
\end{center}
      \end{figure}

Because the WFPC2 and NICMOS images were taken two years apart, it is
important to consider the possibility that the relative astrometry is
corrupted due to motion of CIT~6 with respect to the reference star.  
\citet{bastian88} documented the binary separation of these
sources as 10.4\arcsec$\pm$0.4\arcsec (PA 18\arcdeg$\pm$1\arcdeg) 
at epoch 1955.3 using Palomar Sky Survey (red) plates. The red WFPC2 data
from 1996.3 shows a separation of 11.14\arcsec$\pm$0.02\arcsec
(PA -20.3\arcdeg$\pm$0.1\arcdeg).  
According to these results, the proper motion of CIT~6 
with respect to the reference star is 0.021\arcsec$\pm$0.014\arcsec yr$^{-1}$.
This implies a 0.040\arcsec$\pm$0.027\arcsec shift between the WFPC2 and
NICMOS observations.  This is at the same level as the error on the K band
centroid measurement ($\pm$0.035\arcsec).  Importantly, if one were to
apply this correction it would cause the infrared centroids in 
Figure\,\ref{fig:astrometry} to shift {\em away} from the visible centroid 
towards the North-East.  

\begin{table}[t]
\begin{center}
\scriptsize
\caption{Multi-wavelength Relative Astrometry of CIT~6
\label{table:astrometry}}

\begin{tabular}{c|c|c|lll|lll|l}
\tableline\tableline
Wavelength & Instrument & Filter & \multicolumn{6}{c}{Position Relative to Reference Star} & Comments \\
($\mu$m)   & & & \multicolumn{3}{c}{South Component} & \multicolumn{3}{c}{North Component} & \\
           & & & East (\arcsec) & North (\arcsec) & $\Delta\Theta$ (\arcsec) &
 East (\arcsec) & North (\arcsec) & $\Delta\Theta$ (\arcsec) & \\
\tableline
0.429 & WFPC2 & F439W & -3.870 & -10.516 & 0.015 &  & & & Slight saturation \\
0.525 & WFPC2 & F555W & -3.886 & -10.516 & 0.020 & -3.845 & -10.326 & 0.020 
& Saturation \\
0.674 & WFPC2 & F675W & -3.885 & -10.513 & 0.010 & -3.849 & -10.327 & 0.010 
& No Problems \\
1.129 & NICMOS & F110W & & & & -3.836 & -10.373 & 0.015 & No Problems\tablenotemark{a} \\
1.607 & NICMOS & F160W & & & & -3.827 & -10.374 & 0.022 
& Saturation\tablenotemark{a}~\tablenotemark{b} \\
2.218 & NICMOS & F222M & & & & -3.806 & -10.353 & 0.035
& February 1998; Saturation\tablenotemark{b} \\
2.218 & NICMOS & F222M & & & & -3.809 & -10.357 & 0.035
& May 1998; Saturation\tablenotemark{b} \\
\tableline
\end{tabular}
\tablenotetext{a}{CIT~6 and reference star were not on the same frame and
required the use of a Small Angle Maneuver}
\tablenotetext{b}{The extrapolated intersection of the diffraction spikes
was used to estimate location of CIT~6 centroid due to excessive 
saturation.}
\end{center}
\end{table}

\section{Discussion}

Because the northern component of the Keck 2.2\micron ~images (see
Figure\,\ref{fig:cit6}) dominates the flux, the centroid of the
NICMOS images should lie close to this location.
Based on the HST relative 
astrometry, it appears that the northern component in the
IR images can be associated with the northern component in the red WFPC2
images, although 
they do not appear to be exactly co-spatial.  Interestingly, this places
the faint, southern IR component approximately halfway between the two
visible components.  Note that the small proper motion discussed in 
\S\ref{section:motion} would cause the IR centroids to appear further to the 
North-East, away from the average position of the visible components, 
strengthening the above interpretation.
One interpretation of the two visible components
is that they 
arise from a
bipolar reflection nebula with low density poles allowing scattered
light to escape the dust shell.  This is supported by
the linear polarization of the red emission, with a
PA of $\sim$75\arcdeg \citep{trammell94}.
The dominance of the northern component
in the IR indicates that the northern pole is inclined towards us.  In
this scenario, the faint, southern IR component could be thermal
emission from the central ``disk'' of material which is responsible for
channeling the stellar flux towards the bipolar lobes.

Unfortunately, this appealing model does not account for the blue WFPC2
observations.  In fact, the southern component is brighter in all
the WFPC2 images.  However, the source of blue light is likely to be
different from the source of red emission
\citep{alksnis88}.  
This contention is most
securely based on multi-wavelength photometry which shows that the red
and infrared flux vary at the $\sim$626~day period of the carbon star,
while the blue flux level is roughly constant \citep{alksnis95}.  
In addition, the blue
light has a polarization angle ($\sim$0\arcdeg) 
roughly orthogonal to that seen in the
red and infrared \citep{cohen82,trammell94}.

One scenario able to reconcile these disparate observations is to propose
that the blue lobe is not being illuminated the carbon star, but by a blue
companion which lies in the vicinity of the southern lobe.  
Although it may appear somewhat ad-hoc, this model is able to 
qualitatively explain the odd cometary shape of the bluest WFPC2 image
and the different blue-red polarization angles.  
The apparent 
location of the southern (blue) HST component would be largely defined by 
the mass-loss profile of the carbon star, while the angle of linear 
polarization 
in the blue would be set by the separation vector between the companion and the
scattering lobe.  Note that the polarization direction of the H$_\alpha$ 
emission
observed by \citet{trammell94} matches the blue polarization, not the red.  
This is consistent with the idea that the blue companion is producing 
the line emission which is scattering into our line of sight along 
the same trajectories as the blue continuum.

\subsection{A Dusty Disk?}

\begin{figure}
\begin{center}
\includegraphics[width=4in]{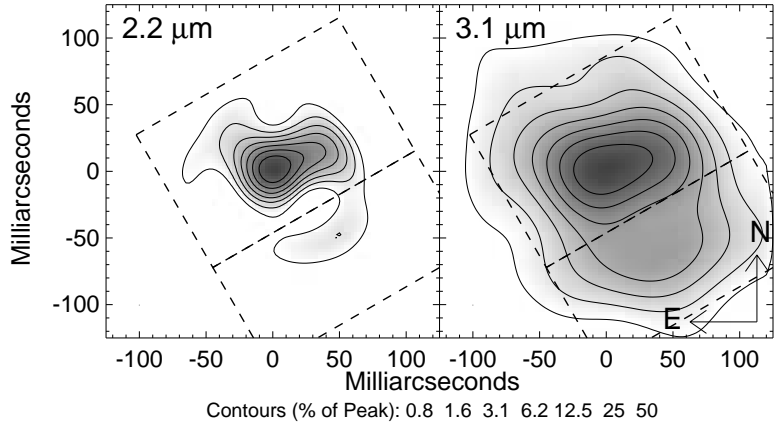} 
      \caption{This figure shows coadded images of CIT~6 at  2.2\micron ~({\em left panel}) 
and 3.1\micron ~({\em right panel}).  The dashed boxes define the northern
 and southern component
for the color temperature analysis in \S\ref{section:temperature}.
Each contour level represents a factor of two in
surface brightness, and corresponds to 0.78\%, 1.56\%, 3.13\%, 6.25\%, 12.5\%, 25\%, and 50\% of the peak intensity.
\label{fig:cit6sum}}
\end{center}
      \end{figure}

\begin{figure}
\begin{center}
\includegraphics[width=4in]{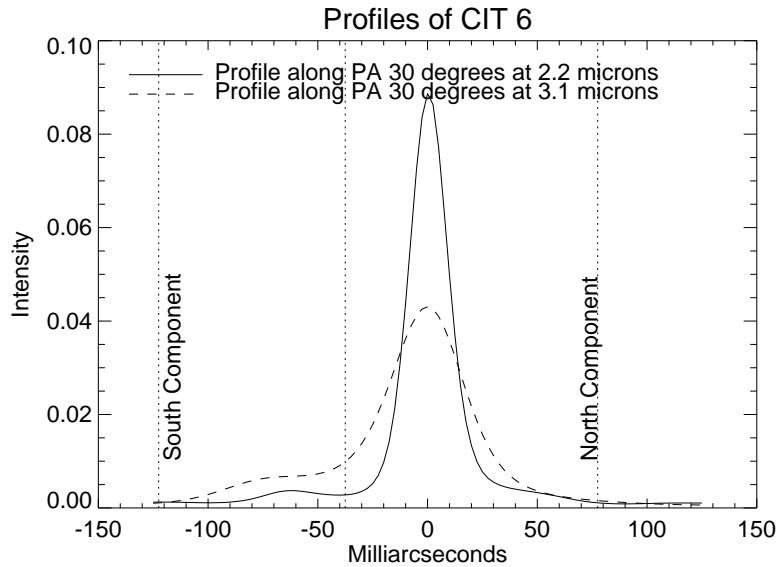} 
      \caption{Profiles of CIT 6.  The coadded images of CIT~6 have been projected onto PA
30\arcdeg ~to isolate contributions from the northern and southern components, and the
curves have been normalized so that the total flux is equal to unity at both
wavelengths.  The dotted lines mark off the areas defined in
\S\ref{section:temperature}.
\label{fig:profiles}}
\end{center}
      \end{figure}

Generally, scattered photospheric light should have a higher color
temperature than thermal emission from dust.  Hence, the above model
can be crudely tested by calculating the color temperature of the
northern and southern IR components.  Towards this goal, all the epochs of
Keck data have been averaged together, where image registration was
estimated by maximizing the cross-correlation; these coadded
images appear in Figure\,\ref{fig:cit6sum}.  It is clear that the
southern component has relatively more flux (compared to the northern
component) at 3.1\micron~than at 2.2\micron, consistent with the
southern component being redder.  These images have been projected
onto an axis at PA 30\arcdeg, and the summed profiles appear in
Figure\,\ref{fig:profiles}.  Note that the larger FWHM of the 
3.1\micron~peak
is due to the lower spatial resolution, and the profiles have been
normalized by the total flux under each curve 
rather than the peak intensity.

To minimize bias due to the wavelength-dependent spatial
resolution, two photometric apertures have been defined to include
most of the flux in the northern and southern IR components.  These
apertures appear in Figure\,\ref{fig:cit6sum} and their projected
boundaries along PA 30\arcdeg are marked on
Figure\,\ref{fig:profiles}.  At 2.2\micron ~(3.1\micron), the southern
aperture contains $\sim$8\% ($\sim$18\%) of the total flux in the two
apertures.  \citet{taranova99} report K and L band magnitudes for
CIT~6 during the period of these Keck observations, with a mean K mag
($\lambda_0 = 2.2\micron$) of 1.83, and L mag ($\lambda_0 =
3.5\micron$) of -1.10.  We linearly interpolate (in 
wavelength--magnitude space)
to arrive at an approximate 3.1\micron~magnitude of 0.22.  Using these
average fluxes we can now estimate the average color temperature of
the northern and southern components separately.

The color temperatures based on the 2.2\micron~and 3.1\micron~images
are 870~K and 600~K for the northern and southern components
respectively.  No color corrections were applied when converting
magnitudes to physical flux units, but any corrections will modify the
component temperature estimates in the same way.  As a comparison, the
mean color temperature based on J and K-band magnitudes is
$\sim$1000~K, according to \citet{taranova99}.  
Although the computation above is clearly an approximation,
the southern IR component has been found to be demonstrably redder than
the northern one.

However, the lower color temperature of the southern component does
not guarantee that this is thermal emission from dust.  The relative
contributions of reddened scattered light and thermal emission are
difficult to untangle in optically thick envelopes
\citep[e.g.,][]{weigelt98a,monnier99a,tuthill2000b}.
The fact that the southern IR component is
located roughly between the visible lobes and has the lowest color
temperature, is certainly consistent with it being 
a dusty ``disk'' of material
surrounding the carbon star.

\label{section:temperature}

\subsection{Time-variability}
The dominant (northern) IR component clearly changes shape on the time 
scale of years, as can be seen from examination of Figure\,\ref{fig:cit6}.
Most notably, the northern IR component has evolved from a nearly symmetric
elongation in 1997 to a very asymmetric structure extending
West-Northwest.  These changes are observed at both wavelengths and
show a consistent evolution in time.
Under the above hypothesis that
this component represents scattered light or thermal emission for
warmed dust, 
this time evolution can be
understood if dust formation closer to the star is changing the
illumination pattern.  
These observations represent the first direct detection of 
evolution of the inner dust shell of CIT 6.
The inner dust formation zone around another nearby
carbon star, IRC+10216, has also 
been observed to be quite chaotic and
dynamic, evolving on similar time scales \citep{tuthill2000a}.
However in IRC~+10216, discrete dust blobs are observed to move away from
central souce, while only slight shifts in the emission pattern are seen for
CIT~6.  This difference can be understood 
if the central star of CIT~6 is more obscured 
than for IRC~+10216, and also by considering that CIT~6 is thought to be
be $\sim$4 times more distant than IRC~+10216.

If illuminated by scattered light,
this changing morphology of the northern IR component could explain the
observed changes in polarization angle of the red and infrared light.
Since 1970, the polarization angle has varied by $\pm$20\arcdeg,
perhaps related to the opening angle of the northern lobe, as can
be seen by the polarization vectors placed on Figure\,\ref{fig:wfpc2}.  As the
center of scattered light changes in this lobe, the corresponding
polarization angle would as well.  Coordinated spectropolarimetry and
high spatial resolution imaging could test this hypothesis further.

\subsection{Alternate models}
Unfortunately, these new data do not preclude other interpretations.
Alternate possible dust shell geometries include the northern component
being the star itself (or a binary), with the southern component arising from
scattered radiation.  As for the case of carbon star IRC~+10216, it
is not obvious if the star itself is visible in existing near-infrared
images, or if it remains obscured
\citep[see discussions in][]{haniff98,weigelt98a,tuthill2000b}.
Yet another possibility is that 
the elongated and time-variable morphology of the northern
IR component could be due to a close binary. 

A direct measure
of the polarization percentage of both IR components would more
definitively determine which (if either) are due to scattered
radiation, although such observations are technically difficult.
Alternate theories for the blue excess, either due to a mistake in the
spectral type identification \citep{trammell94} or circumstellar shocks
related to mass-loss \citep{cohen82}, do not seem able to explain either the
odd cometary appearance or the polarization properties without
positing an additional bipolar nebula at right-angles to the first.
Given the IR morphology, the wide binary hypothesis seems the best
hypothesis of those presently considered for explaining the 
peculiar properties of the blue light, but stronger evidence is
needed to confirm it.

\section{Conclusions}

The dust shell of carbon star CIT~6 has been examined with
unprecedented spatial resolution ($\simle$50\,mas) using aperture
masking at the Keck-I telescope.  Serendipitously, a nearby ($\sim$11\arcsec) 
companion allowed high precision relative astrometry from 
recent HST WFPC2 and NICMOS images.
Multi-wavelength
analysis suggests that CIT~6 is surrounded by a bipolar dust envelope
with a complex morphology, consisting of at least two components.
The dominant compact feature in the near-infrared is tentatively
identified as the northern scattering lobe of the reflection nebula,
although the possibility that it is the carbon star itself can not be
ruled out.
In addition, a fainter IR component with very red colors has been
detected between the two visible light components, perhaps due to
thermal emission from the central ``disk'' of the bipolar dust
envelope.
Excess blue emission in the southern optical lobe is
ascribed to an obscured hot companion, responsible for the strong
wavelength-dependence of the (visible) polarization properties and
``cometary'' scattering feature in the blue WFPC2 image.
Under this model, 
changes in the red and infrared polarization angle should be
linked to observed variations in the near-IR emission morphology.

Other recent high resolution observations of evolved stars have
also revealed strong deviations form spherically symmetry.  Red 
supergiant VY~CMa \citep{monnier99a} and carbon star IRC~+10216
\citep{haniff98,weigelt98a,tuthill2000b} also show
evidence for bipolar nebulae on large scales (as seen in scattered 
light) and complicated dust shells in the innermost regions (from
IR observations).  The dominant physical mechanism for breaking
spherical symmetry and producing the bipolar mass-loss profile 
remains unidentified.


\acknowledgments

The authors would like to recognize Dean Hines, Gary Schmidt, and
Susan Trammell for sharing results of their independent studies in
advance of publication and for stimulating conversations regarding the
CIT~6 dust shell.  In addition, we thank Eddie Bergeron and Glenn
Schneider for advice regarding relative astrometry with HST.  The
image reconstructions presented here were produced by the
maximum-entropy mapping program ``VLBMEM,'' written by Devinder Sivia.
This research has made use of the SIMBAD database, operated at CDS,
Strasbourg, France, and NASA's Astrophysics Data System Abstract
Service.  The data presented herein were obtained at the W.M. Keck
Observatory, which is operated as a scientific partnership among the
California Institute of Technology, the University of California and
the National Aeronautics and Space Administration.  The Observatory
was made possible by the generous financial support of the W.M. Keck
Foundation.  These results are based also on observations made with
the NASA/ESA Hubble Space Telescope, obtained from the data archive at
the Space Telescope Science Institute. STScI is operated by the
Association of Universities for Research in Astronomy, Inc. under NASA
contract NAS 5-26555.  This work is a part of a long-standing
interferometry program at U.C. Berkeley, supported by the National
Science Foundation (Grants AST-9321289 and AST-9731625) and NASA.
JDM was partially supported by a Postdoctoral Fellowship from the 
Harvard-Smithsonian Center for Astrophysics.

\bibliographystyle{apj}
\bibliography{apj-jour,CIT_6,Thesis}



\clearpage








\clearpage

\end{document}